\documentclass[aps,prl,twocolumn,showpacs,final]{revtex4}
\usepackage{graphics}

\bibliography{20}
\begin{document}
\title{Total absorption of an electromagnetic wave by an overdense plasma}
\author{Yury P. Bliokh}
\affiliation{Physics Department, Technion-Israel Institute of Technology, 32000 Haifa, Israel}
\author{Joshua Felsteiner}
\affiliation{Physics Department, Technion-Israel Institute of Technology, 32000 Haifa, Israel}
\author{Yakov Z. Slutsker}
\affiliation{Physics Department, Technion-Israel Institute of Technology, 32000 Haifa, Israel}
\begin{abstract}
We show both theoretically and experimentally that an electromagnetic wave can be totally absorbed by an overdense plasma when a subwavelength diffraction grating is placed in front of the plasma surface. The absorption is due to dissipation of surface plasma waves (plasmons-polaritons) that have been resonantly excited by the evanescent component of the diffracted electromagnetic wave. The developed theoretical model allows one to determine the conditions for the total absorption.
\end{abstract}
\pacs{52.40.Db, 73.20.Mf, 42.25.Fx}
\maketitle

Media with negative permittivity $\varepsilon<0$ have recently attracted enhanced attention. Periodically-inhomogeneous (perforated or corrugated) optically-thick metal films are anomalously transparent for light at certain resonance frequencies for which the metal permittivity is negative \cite{Ebbesen}. Media with both negative permittivity $\varepsilon<0$ and negative permeability $\mu<0$, the so-called left-handed materials (LHM) \cite{Veselago,Pendry1}, possess unusual properties, in particular the possibility of creating an ``ideal lens'' \cite{Pendry2} with subwavelength resolution. A common property for  metals with $\varepsilon<0$ and for LHM is that the boundary between such media and vacuum supports propagation of surface waves, the so-called plasmons-polaritons \cite{Ritche, Rupin}. In fact, the surface of a medium with $\varepsilon<0$ behaves as a surface wave resonator (SWR). The resonators on both sides of a slab made of an $\varepsilon<0$ material are coupled by their exponentially vanishing fields. When one of the resonators is excited by an external field, then, due to the coupling, the second resonator is also excited. It means that the external field leaks through the slab. When the resonators' Q-factors are large, the amplitude of oscillations in the second resonator is the same as in the first one and the slab transparency reaches 100\%.

A propagating incident electromagnetic wave cannot excite the SWR. Only non-propagating, evanescent waves for which $\omega/c<k_\perp$ ($\omega$ is the wave frequency and $k_\perp$ is the tangential component of the wavenumber $\vec{k}$) can play the role of a pump mode and excite the resonator. Exactly such waves excite the SWR of an LHM slab and enable restoration with subwavelength resolution of the image of the wave source at the opposite side of the slab. There are two ways to excite the SWR on a metal surface by an incident electromagnetic wave. It is possible to modify the metal surface so that the modified surface waves can be in resonance with the electromagnetic wave. The second way consists of a transformation (partial or complete) of the propagating incident wave into an evanescent one. The first way is realized by forming periodical inhomogeneities on the metal surface. The second way can be realized by total internal reflection at a near-boundary inhomogeneity of $\varepsilon$ \cite{Otto, Aliev} or by using a diffraction grating \cite{Bliokh}.

The electromagnetic properties of a metal in such a frequency range where $\varepsilon<0$ are similar to those in plasma. The plasma permittivity $\varepsilon=1-\omega^2_p/\omega^2$ is negative when $\omega<\omega_p$ (here $\omega_p$ is the plasma frequency). In spite of such similarity of metal and plasma, all experimental investigations have been carried out with metal films only and with electromagnetic waves in the visible or near-infrared frequency bands. In this Letter we report for the first time on an experimental observation of strong absorption of microwaves by an overdense plasma due to excitation of surface plasma waves.

The application of diffraction gratings for partial transformation of incident propagating electromagnetic waves into evanescent ones is the most convenient for plasma experiments. Note that forming a special profile of the plasma density, as suggested in Ref.~\cite{Aliev}, is considerably more difficult. It seems that by placing two diffraction gratings on the opposite sides of a plasma layer it is possible to realize electromagnetic wave transmission through an overdense plasma with $\omega_p>\omega$ or even $\omega_p\gg\omega$. The transparency of such a ``sandwich'' with a dissipationless plasma inside reaches 100\% \cite{Bliokh}. However, it is practically impossible to realize such a situation experimentally. The point is that the coupling between the SWRs on the two sides of the plasma layer decreases exponentially when one increases the layer thickness. The weaker is the coupling between resonators, the more easily can dissipation destroy this coupling. The resonators are in fact isolated and excitation is not transmitted from one resonator to the other, i.~e. the electromagnetic energy does not pass through the plasma layer. Since it is practically impossible to create a dense plasma in a very thin layer, the transmission of an electromagnetic wave through an overdense plasma is also impossible. Note that the same destructive role of dissipation significantly restricts the subwavelength resolution of the ``ideal lens'' \cite{N-V}.

In practice an incident electromagnetic wave can excite only one SWR on one side of the plasma layer. However, the  resonant interaction of the electromagnetic wave with the plasma surface can manifest itself in a total absence of a reflected wave. The incident wave is completely transformed into a surface wave which dissipates in the plasma. This is the phenomenon of the {\it total absorption} of an electromagnetic wave in the overdense plasma.

The problem of excitation of surface waves  is equivalent to the problem of resonator excitation \cite{Bliokh}. Using this analogy, let us consider the problem of wave transmission through a one-dimensional resonator with semi-transparent walls. The transparency of the two walls will be characterized by transmission coefficients $T_1$ and $T_2$.

Let us first consider the resonator without a pump wave.  The energy $W_r$ stored in the resonator can be expressed through a complex amplitude $E_r$ of field oscillations:
\begin{equation}\label{eq1}
W_r=N{|E_r|^2\over 4\pi},
\end{equation}
where $N$ is the resonator eigenmode norm.
The resonator energy loss is determined by the dissipation in the resonator and by the energy flux through the walls. The dissipative energy loss can be written in the form:
\begin{equation}\label{eq2}
\left.{dW_r\over dt}\right|_{diss}=-2\gamma W_r.
\end{equation}
Here $\gamma$ is the damping coefficient of the oscillations  in the medium, $|E_r(t)|\propto\exp(-\gamma t)$.

The energy fluxes $P_{1,2}$ of the waves leaking through the walls of the resonator are equal to:
\begin{equation}\label{eq3}
P_{1,2}=v_g{|E_{1,2}|^2\over 4\pi},
\end{equation}
where $v_g$ is the wave group velocity and $E_i =\sqrt{T_i}E_r$ are the wave amplitudes. So the energy conservation law has the form:
\begin{equation}\label{eq5}
{d\over dt}|E_r|^2=-|E_r|^2\left[2\gamma+{v_g\over N}(T_1+T_2)\right].
\end{equation}

Let us make use of the analogy between a resonator and an oscillator and consider an oscillator with eigenfrequency $\omega_r$ and dissipation coefficient $\alpha$ that coincide with the corresponding characteristics of the resonator. It follows from the oscillator equation
\begin{equation}\label{eq6}
{d^2E_r\over dt^2}+\omega_r^2E_r+\alpha{dE_r\over dt}=0
\end{equation}
and Eq.~(\ref{eq5}) that 
\begin{equation}\label{eq8}
\alpha=2\gamma+{v_g\over N}(T_1+T_2).
\end{equation}

Let us now consider  the problem of the resonator excitation by a wave which falls on the left resonator wall. Let the incident wave amplitude be $E_0$. The amplitude of the wave that penetrates the resonator is $\sqrt{T_1}E_0$. The field of this wave appears as an external force in the oscillator equation:
\begin{equation}\label{eq9}
{d^2E_r\over dt^2}+\omega_r^2E_r+\alpha{dE_r\over dt}=q\sqrt{T_1}E_0e^{-i\omega t},
\end{equation}
where $q$ is some complex coefficient. One cannot put $q=1$ because not only the phase of the field that penetrates the resonator is unknown, but also the efficiency of the resonator excitation by this field is unknown. This efficiency is determined by the projection of the penetrating field on the field of the resonator eigenmode.

Using the solution of Eq.~(\ref{eq9}), one can write the following expression for the amplitude $|E_T|$ of the wave transmitted through the resonator:
\begin{equation}\label{eq11}
|E_T|=\sqrt{T_2}|E_r|=|E_0|\left|{q\sqrt{T_1T_2}\over\omega_r^2-\omega^2-i\alpha\omega}\right|.
\end{equation}

There are many examples that in systems analogous to the one under consideration when dissipation is absent $(\gamma=0)$, the transmission coefficient at a resonance frequency is equal to unity when the system possesses a mirror symmetry. In quantum mechanics it is the resonant tunnelling through a symmetric two-humped potential barrier \cite{Bohm}. Localized states in disordered media and resonant transmission which is associated with these states, present another example of a symmetric system \cite{Localization}. Examples that are closer to the problem under consideration have been described in Refs.~\cite{Fan, Yariv1, Yariv2}. It is reasonable to suppose that our system also possesses the same property, namely the transmission coefficient $T=|E_T|^2/|E_0|^2$ reaches unity (and the reflection coefficient $R$ vanishes) at the resonance frequency when dissipation is absent, $\gamma=0$, and the system is symmetric, i.~e. $T_1=T_2$. It follows from the condition $T=1$ and the definition (\ref{eq8}) for $\alpha$ that
\begin{equation}\label{eq13}
|q|={2\omega_r v_g\over N}.
\end{equation}

In order to calculate the reflection coefficient $R$, let us write the energy conservation law:
\begin{equation}\label{eq16}
v_g{|E_0|^2\over 4\pi}(1-R)=v_g{|E_T|^2\over 4\pi}+2\gamma N{|E_r|^2\over 4\pi}.
\end{equation}
It follows from Eq.~(\ref{eq16}) that
\begin{equation}\label{eq17}
1-R=\left({2\omega_rv_g\over N}\right)^2{T_1\left(T_2+2\gamma N/v_g\right)\over \left|\omega_r^2-\omega^2-i\alpha\omega\right|^2}.
\end{equation}
The reflection is minimal at the resonance frequency and  we will put further $\omega=\omega_r$. In the problem under  consideration the second wall of the resonator, namely the plasma body, is opaque, $T_2=0$. In spite of the mirror symmetry breakdown, the reflection coefficient vanishes under a certain condition.  Indeed, by setting $T_2=0$ in Eq.~(\ref{eq17}), one can obtain:
\begin{equation}\label{eq18}
R=1-{4u\over(1+u)^2},\,\,{\rm where}\,\, u={2\gamma N\over T_1 v_g}.
\end{equation} 
Thus, the reflection coefficient is equal to zero when the dissipation coefficient $\gamma$ is equal to a certain critical  value $\gamma_c=T_1v_g/(2N)$. The analogous phenomenon in symmetric systems is known as {\it critical coupling} \cite{Yariv1, Yariv2}.

The phenomenological parameter $T_1$ which appears in the developed theoretical model, can be estimated in the following way. Let $\eta$ be the transformation ratio of the propagating wave into the evanescent one on the grating. Then $T_1^{1/2}\sim\eta\exp(-k_vd)$, where $d$ is the distance between the grating and the plasma boundary, and $k_v$ is the spatial decrement of the evanescent wave amplitude in vacuum. This parameter plays in our theory the same role as the coupling coefficient in \cite{Yariv1, Yariv2}.

The results of the above simple theory are in good agreement with the solution of a complete system of equations which describe the diffraction of the incident wave in the grating and the interaction of the diffracted fields with the dissipative overdense plasma.  It was assumed in the calculations that the grating is subwavelength, i.~e. the grating wavenumber $k_g>\omega/c$. It has been found that the reflection coefficient as a function of $\omega$ is minimal, $R(\omega)=R_{min}$, when the wavenumber $k_\perp=k_x+k_g$ (where $k_x$ is the projection of the wavenumber $\vec{k}$ of the incident wave on the grating plane) coinsides with the wavenumber $k_s$ of the plasma surface wave which propagates in the $x$ direction:
\begin{equation}\label{eq19}
k_s=k_0\sqrt{\left(\omega_p^2-\omega^2\right)/\left(\omega_p^2-2\omega^2\right)},
\end{equation}
where $k_0=\omega/c$. The fields of the surface wave decay exponentially on both sides of the plasma boundary. This surface wave forms the resonator considered above.

\begin{figure}[htb]
\centering \scalebox{0.35}{\includegraphics{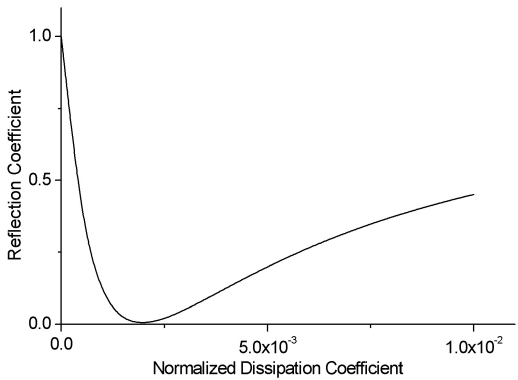}}
\caption{\label{Fig1}The reflection coefficient $R_{min}$ versus the normalized dissipation coefficient $\Gamma$.}
\end{figure}

As an example, the dependence of $R_{min}$ on the normalized dissipation coefficient $\Gamma=\gamma/\omega_p$  for certain values of $\eta$ and $d$ is presented in Fig.~\ref{Fig1}. The critical coupling effect is clearly visible in this picture. 

In our experiments the ferroinductor coupled plasma discharge \cite{Slutsker} was used as the plasma source. The plasma appears as a result of a gas breakdown due to the electric field surrounding a ferrite core with a large $\mu$, when an rf voltage is applied to the core primary winding. We used a 1~ms pulse of rf with a frequency of 240~kHz. This source produces a dense homogeneous plasma, the density of which can be varied over a large range by varying the gas pressure and/or the power of the plasma source. In such source no magnetic field exists in the plasma volume and at its boundary as well. In the absence of magnetic field, a surface plasma wave can propagate on the plasma boundary.

\begin{figure}[htb]
\centering \scalebox{0.25}{\includegraphics{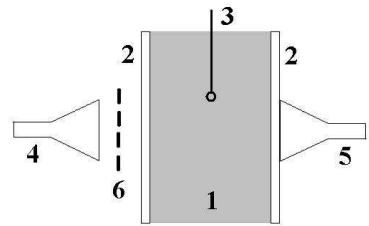}}
\caption{\label{Fig2}Experimental setup. 1 -- plasma layer, 2 -- plastic disk, 3 -- movable probe, 4 -- microwave transmitting antenna, 5 -- microwave receiving antenna, 6 -- diffraction grating.}
\end{figure}

A sketch of the experimental setup is shown in Fig.~\ref{Fig2}.  Two  movable thin plastic plates were located on each side of the core. These plates bordered the volume filled with the plasma and in fact created the homogeneous plasma considered in the theory. To measure the plasma density spatial distribution we used a movable single probe. Indeed the plasma inhomogeneity along the chamber axis did not exceed 5\%. Also in the radial direction the plasma density fall did not exceed 10\% at a distance of 8-9~cm from the chamber axis. Due to the movable plates the thickness of the plasma layer could be varied in the range of 9-20~cm. The plasma density $n_p$ could reach $4.5\cdot10^{12}\,{\rm cm}^{-3}$. The neutral gas pressure $p$ was varied within the range of $2\cdot10^{-4}-2\cdot10^{-3}$~Torr of Xe. The ionization level was varied within the range of 7--30\% for the high and low  pressures respectively. The electron temperature was found to be about 4~eV and the electron collision frequency should reach $3\cdot10^7\,{\rm sec}^{-1}$ at the highest pressure.

Two rectangular microwave horn antennas were placed close to the outer surfaces of the plastic plates. In these plates in front of the horn openings were made windows, which were covered by thin (0.2~mm) teflon films in order to prevent microwave reflection from the plates. One of these horns was connected via a ferrite circulator to a microwave oscillator with wavelength $\lambda=3.2\,{\rm cm}$. This circulator allowed us to separate the input and reflected microwave signals. The other horn was used in order to verify the probe measurements by the microwave cut-off method. 
 
The experiments were carried out with a diffraction grating which consisted of a periodical set of metal strips of 11~mm width and a period of $20$~mm. The distance $d$ between this grating and the plasma boundary could be varied in the range of 3-15~mm. The electromagnetic wave was incident normal to the grating and the plasma surfaces, i.~e. $k_\perp=k_g$. Under such circumstances, we obtain from Eq.~(\ref{eq19}) that the plasma surface wave should be excited at a plasma density of $n_p=2.9\cdot10^{12}{\rm cm}^{-3}$. This density exceeds by 2.6 times the critical density $n_c=1.1\cdot10^{12}\,{\rm cm}^{-3}$ corresponding to $\lambda=3.2$~cm.

\begin{figure}[htb]
\centering \scalebox{0.37}{\includegraphics{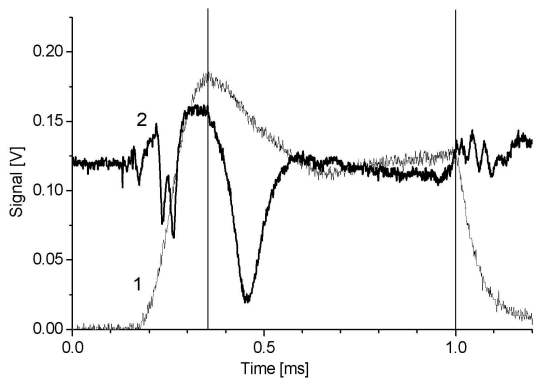}}
\caption{\label{Fig3}Typical scope traces of the probe signal (1) and reflected signal (2). Only the time interval between the vertical lines corresponds to a stationary plasma. }
\end{figure}

Typical waveforms of the reflected wave and probe signals are shown in Fig.~\ref{Fig3}. The time intervals before the first vertical line and after the second one should be excluded from our consideration. The reason is that they correspond either to the initial stage of the gas breakdown and plasma formation  (the first interval) or to the plasma decay after the driving pulse termination (the last interval) while we should deal only with the middle interval which corresponds to a stationary plasma.
\begin{figure}[htb]
\centering \scalebox{0.83}{\includegraphics{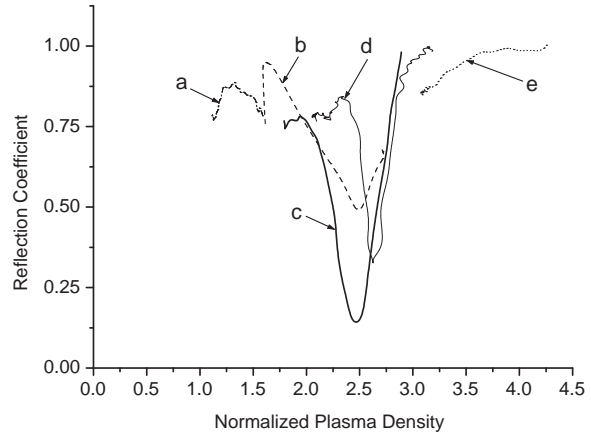}}
\caption{\label{Fig4}Dependencies $R(n_p)$. The plasma density is normalized to the critical density $n_c$. (a)\, $p=1.6\cdot10^{-3}$~Torr, $d=9$~mm; (b)\, $p=3.1\cdot10^{-4}$~Torr, $d=13$~mm; (c)\, $p=6\cdot10^{-4}$~Torr, $d=9$~mm; (d)\, $p=1.6\cdot10^{-3}$~Torr, $d=5$~mm; (e)\, $p=1.6\cdot10^{-3}$~Torr, $d=9$~mm;}
\end{figure}

Several experiments were carried out in order to cover a broad range of plasma densities, $n_c<n_p<4n_c$. Within all this range the reflection coefficient $R$ was close to unity except for a narrow range of $n_p\simeq2.0\,n_c-2.7\,n_c$ where it had a deep minimum at a certain resonance density $n_r$, $R(n_r)=R_{min}$. Indeed in this range the value of $R$ was several fold smaller than the reflection from the grating itself ($R_g\simeq0.8$). The dependencies $R(n_p)$ for various $d$ and $p$ are presented in Fig.~\ref{Fig4}. The value of $n_r$ was found to be only weakly dependent on the parameters $d$ and $p$, and independent of the distance between the plates. On the other hand the value of $R_{min}$ did depend on $d$ and $p$ more strongly. By varying $p$, the minimum value of $R_{min}$  at each $d$ depended on $d$ nonmonotonically. The smallest of all values of $R_{min}$, namely $R_{min}\simeq 0.14$, corresponds to a distance of $d=9\,{\rm mm}$ (for instance $R_{min}\simeq0.3$ for $d=5\,{\rm mm}$ and $R_{min}\simeq0.5$ for $d=13\,{\rm mm}$). It seems likely that the critical coupling takes place when $d$ is close to $9\,{\rm mm}$.

In conclusion we have shown theoretically and experimentally that total absorbtion of electromagnetic waves may be achieved in  overdense plasmas when a properly designed diffraction grating is placed in front of the plasma boundary. This phenomenon is caused by resonant excitation of surface plasma waves, plasmons-polaritons, and their dissipation in the plasma. Such possibility of total microwave absorption  is very attractive for a number of important applications, e.~g. plasma heating and nonreflecting coating.
\begin{acknowledgments}
%This work was supported \ldots
\end{acknowledgments}


\begin{references}
\bibitem{Ebbesen} {T. W. Ebbesen, H. J. Lezec, H. F. Ghaemi, T. Thio, and P. A. Wolff, Nature {\bf 391}, 667 (1998).}
\bibitem{Veselago} { V. G. Veselago,  Sov. Phys. Usp. {\bf 10}, 509 (1968).}
\bibitem{Pendry1} Focus Issue: Negative Refraction and Metamaterials,  Opt. Express {\bf 11}, 639 (2003).
\bibitem{Pendry2}J. B. Pendry, Phys. Rev. Lett. {\bf 85}, 3966
(2000). 
\bibitem{Ritche} R. H. Ritche, Phys. Rev. {\bf 106}, 874 (1957).
\bibitem{Rupin}R. Rupin, J. Phys.: Cond. Mat. {\bf 13}, 1811
(2001).
\bibitem{Otto} A. Otto, Zeitschr. Phys. {\bf 216}, 398 (1968).
\bibitem{Aliev} Yu. M. Aliev, O. M. Gradov, A. Yu. Kyrie, V. M. $\check{\rm C}$ade$\check{\rm z}$, and S. Vukovi\'{c}, Phys. Rev. A {\bf 15}, 2120 (1977).
\bibitem{Bliokh} Yu. P. Bliokh, arXiv:physics/0505078.
\bibitem{N-V} M. Nieto-Vesperinas, J. Opt. Soc. Am. A {\bf 21}, 491 (2004).
\bibitem{Bohm} D. Bohm, Quantum Theory (Prentice-Hall, New York, 1952).
\bibitem{Localization} U. Frisch, C. Froeschle, J.-P. Scheidecker, and P.-L. Sulem,
Phys. Rev. A \textbf{8}, 1416 (1973).
\bibitem{Fan} S. Fan, P. R. Villeneuve, and J. D. Joannopoulos, Phys. Rev. Lett. {\bf 80}, 960 (1998).
\bibitem{Yariv2} A. Yariv, Electronics Lett. {\bf 36}, 321 (2000).
\bibitem{Yariv1} Y. Xu, Y. Li, R. K. Lee, and A. Yariv, Phys. Rev. E {\bf 62}, 7389 (2000).
\bibitem{Slutsker} Yu. P. Bliokh, J. Felsteiner, Ya. Z. Slutsker, and P. M. Vaisberg, Appl. Phys. Lett. {\bf 85}, 1484 (2004).
\end{references}
\end{document}